\documentclass[twocolumn,aps,prb,showpacs,amsmath,superscriptaddress]{revtex4}
\usepackage{amssymb}
\usepackage{mathrsfs}
\usepackage{graphicx}
\usepackage{float}
\usepackage{txfonts}
\begin{document}
\title{Characterization of topological phase transitions via topological properties of transition points}
\author{Linhu Li}
\affiliation{Beijing National Laboratory for Condensed Matter
Physics, Institute of Physics, Chinese Academy of Sciences, Beijing
100190, China}
\author{Shu Chen}
\email{schen@aphy.iphy.ac.cn} \affiliation{Beijing National
Laboratory for Condensed Matter Physics, Institute of Physics,
Chinese Academy of Sciences, Beijing 100190, China}
\affiliation{Collaborative Innovation Center of Quantum Matter, Beijing, China}
\begin{abstract}
  We study topological properties of phase transition points of topological quantum phase transitions by assigning a topological invariant defined  on a closed circle or surface surrounding the phase transition point in the parameter space of momentum and transition driving parameter. By applying our scheme to the  Su-Schrieffer-Heeger model and Haldane model, we demonstrate that the topological phase transition can be well characterized by the defined topological invariant of the transition point, which reflects the change of topological invariants of topologically different phases across the phase transition point.
\end{abstract}

\maketitle
\date{today}

\section{ Introduction}
Conventional continuous quantum phase transitions (QPTs) are driven by pure quantum fluctuation effects due to the change of external parameters and
generally described in terms of the spontaneous symmetry breaking and order parameters of the ground state \cite{Sachdev}. On the contrary, topological QPTs involve the change
of ground-state topological properties and accompany no symmetry breaking \cite{TI-RMP1,TI-RMP2,Volvik}. Different topological states are classified by topological quantum
numbers, which take discrete numbers, in contrast to order parameters used in conventional QPTs to distinguish various phases, which generally take continuous values.
Conventionally, continuous QPTs can be classified into different order QPTs  by  singularity properties of the ground-state energy at the phase transition point
(or critical point for $n \geq 2$), i.e., $n$th order QPTs are characterized by discontinuities in the $n$th derivative of the ground-state energy.

As the singularity of ground-state energy plays an important role in determining universal properties around the critical point of the QPT, however it can not distinguish whether the phase transition is a topological QPT or a conventional one within the Landau-Ginzburg paradigm \cite{Sachdev}. Beyond the traditional energy criterion, a QPT can be also witnessed by qualitative changes of physical quantities related to the ground-state wavefunctions, e.g., the Berry phase \cite{Zhu,Pachos}, quantum fidelity and the fidelity susceptibility \cite{Fidelity,ZhouHQ,GuSJ,Chen},
and the quantum geometric tensor \cite{Zanardi,Ma,Ryu}. Although these approaches have shed light on our understanding of QPTs from the geometric aspect of the ground-state manifold, one can not identify a QPT to be a topological or trivial one solely from the singularity of particular physical quantities at the phase transition point unless additional quantities related to the topological invariant are calculated.
An interesting question arising here is whether we can characterize a QPT is topological or conventional phase transition from the property of the phase transition point?

To answer the question, let us recall that a topological QPT distinguished from a trivial one is manifested by the change of topological invariant, instead of symmetry breaking, across the transition point. While the topological invariant, e.g., the quantized Berry phase for one-dimensional (1D) topological systems \cite{Shen-book} or Chern number for two-dimensional (2D) quantum Hall systems \cite{TKNN}, is well defined for a gapped phase apart from the QPT point and characterizes the global geometrical property of the Bloch band, it fails to work at the gapless critical point. To overcome the difficulty, in this work we propose an alternative definition for the topological invariant, which is not defined on the momentum space at the transition point, but via a closed detour path surrounding the critical point on the parameter space spanning by both the momentum and the transition driving parameter. By applying this idea to the 1D and 2D topological systems, e.g., the celebrated Su-Schrieffer-Heeger (SSH) model and Haldane model, we demonstrate these topological invariants taking nontrivial quantized numbers for topological QPTs, but some non-universal numbers or zero number for conventional QPTs. Our results suggest that we can judge topological or trivial QPTs from topological properties of the phase transition points.

\section{Models and results}
\subsection{1D topological models}
\begin{figure}
\includegraphics[width=0.8\linewidth]{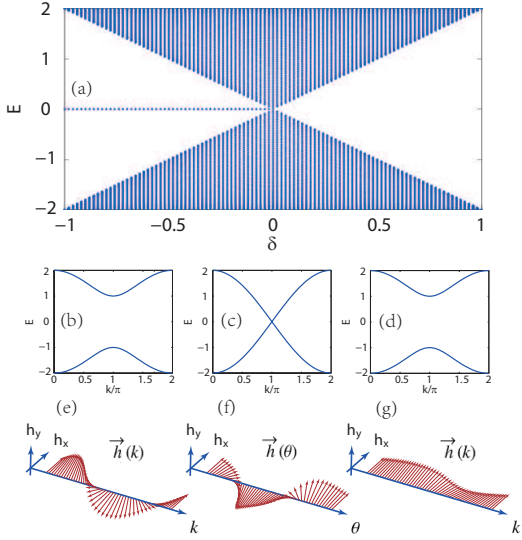}
\caption{ (Color online) (a) The spectrum of the SSH model versus $\delta$ under OBC.
(b)-(d) The spectrum versus $k$ under PBC, with (b) $\delta=-0.5$,
(c) $\delta=0$ and (d) $\delta=0.5$. (e) and (g) show the winding of
$h(k)$ across the Brillouin zone, corresponding to (b) and (d)
respectively. (f) shows the winding of $h(\theta)$ as $\theta$
varies a period. The arrows in (e)-(g) show the direction of the
Hamiltonian $h(k)$ and $h(\theta)$.} \label{Fig1}
\end{figure}

We begin our discussion
with one of the simplest 1D topological systems, the SSH model \cite{SSH},
which can be described by the Hamiltonian:
\begin{eqnarray}
H=\sum_i [(t+\delta)\hat{c}^{\dagger}_{A,i}\hat{c}_{B,i}+(t-\delta
)\hat{c}^{\dagger}_{A,i+1}\hat{c}_{B,i}]+h.c.,
\end{eqnarray}
where $\hat{c}^{\dagger}_{A(B),i}$ is the creation operator of fermion on
$i$-th A (or B) sublattice. This model has two sites in a unit cell,
the hopping amplitude in the unit cell is $t+\delta$ and that
between two unit cells is $t-\delta$. For convenience, $t=1$ is taken as the energy unit. After the Fourier
transformation $\hat{c}_{s,j}=\frac{1}{\sqrt{L}}\sum_k e^{ikj}\hat{c}_{k,s}$ with $s=A (B)$,
the Hamiltonian can be written as
\begin{eqnarray}
H = \psi^{\dagger}_{k} h(k) \psi_{k} ,
\end{eqnarray}
where
$\psi^{\dagger}_{k}=(\hat{c}^{\dagger}_{k,A},\hat{c}^{\dagger}_{k,B})$
and $h(k)=h_x\sigma_x+h_y\sigma_y$ with $\sigma$ the Pauli matrix
acting on the vector $ \psi_{k}$,
$h_x=(1+\delta)+(1-\delta)\cos{k}$, $h_y=(1-\delta)\sin{k}$. It is
well known that this model belongs to the BDI class  according to
the standard topological classification \cite{tenfold} and has two topologically distinct phases for
$\delta>0$ and $\delta<0$ with the phase transition point at
$\delta=0$, where the gap closes at $k=\pi$. Under the open boundary
condition (OBC), these two phases can be distinguished by the
presence and absence of degenerate zero-mode edge states
\cite{Ryu2}, as shown in Fig.\ref{Fig1}(a).

While the spectrum under periodic boundary
condition (PBC) shows a similar structure for $\delta<0$ or $\delta>0$, as displayed in Fig.\ref{Fig1}(b)-(d),
the topological property of the distinct phase can be characterized by the
Zak phase \cite{Zak,Niu}, i.e., the Berry phase across the the Brillouin zone, which is defined
as
\begin{eqnarray}
\gamma=i\int_{-\pi}^{\pi}dk\langle\varphi(k)|\partial_k|\varphi(k)\rangle,
\end{eqnarray}
with $\varphi(k)$ the eigenstate of the occupied Bloch band. While the topological phase
is characterized by $\gamma=\pi$ for $\delta<0$, the trivial phase corresponds to
$\gamma=0$ for $\delta>0$. The geometrical meaning of the Zak
phase can be understood as the winding angle of $h(k)$ as $k$ varies
across the Brillouin zone \cite{Montambaux}, as shown in Fig.\ref{Fig1}(e) and (g).
For the topological nontrivial case with $\delta=-0.5$, the
direction of $h(k)$ winds an angle of $2\pi$, whereas for the trivial
case with $\delta=0.5$ the winding angle is zero. However, when
$\delta=0$, the two bands are degenerate at $k=\pi$
(Fig.\ref{Fig1}(c)), and the Zak phase is ill-defined.

To describe the topological property of state at the phase transition point, here we defined
the Berry phase on a circle around the gap closing point in the
parameter space of $k$ and $\delta$. Introducing $\theta$ as the
varying angle and $A$ as radius of the circle, we have
$k=A\sin{\theta}+\pi$ and $\delta=A\cos{\theta}$, hence the
Hamiltonian around the circle can be represented as $h(\theta)$, and the Berry phase is
defined as
\begin{eqnarray}
\gamma_d&=&i\int_{-\pi}^{\pi}d\theta\langle\varphi(\theta)|\partial_\theta|\varphi(\theta)\rangle. \label{gammad}
\end{eqnarray} After some algebras, one can
obtain $\gamma_d=\pi$, which corresponds to the topological phase
transition at $\delta=0$. In Fig.\ref{Fig1}(f), we also show the winding
of $h(\theta)$ as $\theta$ varies a period, giving rise to a winding angle of $2\pi$.

As a comparison, we next consider a topologically trivial two-band model with alternating on-site potentials, described by the Hamiltonian:
\begin{equation}
H= \sum_i
t[\hat{c}^{\dagger}_{A,i}\hat{c}_{B,i}+\hat{c}^{\dagger}_{A,i+1}\hat{c}_{B,i}+h.c.]
%\nonumber\\
+\mu(\hat{c}^{\dagger}_{A,i}\hat{c}_{A,i}-\hat{c}^{\dagger}_{B,i}\hat{c}_{B,i}).
\end{equation}
This model has alternating chemical potential $\mu$ and $-\mu$ for site A and B,
and has a similar spectrum as the SSH model with a phase transition occurring at $\mu=0$.
However, as $\mu$ breaks the inversion
symmetry, the Berry phase of each band is no longer quantized and no
degenerate edge states emerge under the OBC.
Similarly, we can also calculate the Berry phase around the gap closing point at $\mu=0$ and $k=\pi$.
The numerical result shows that $\gamma_d$ is not a quantized invariant and is
associated with the radius $A$ of the integral path.
%which indicates
%that although the gap closed and reopened at $\mu=0$, this is not a
%topological phase transition.
On the contrary, the $\gamma_d$ for
the SSH model is always $\pi$ regardless of the value of $A$, which suggests that $\gamma_d$ is a topological invariant. This difference means that we can judge whether the QPT is a topological phase transition from values of the Berry phase $\gamma_d$ around the critical point.

\begin{figure}
\includegraphics[width=0.8\linewidth]{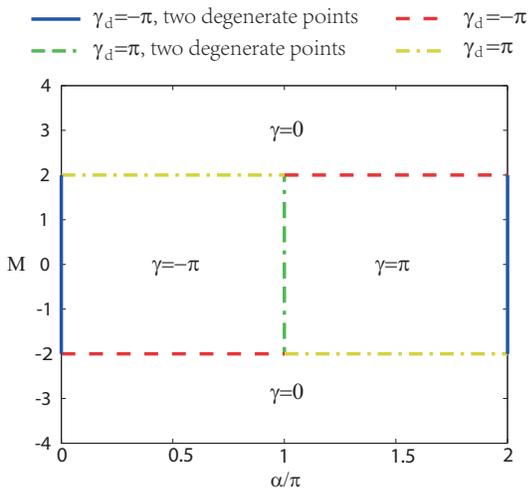}
\caption{(Color online) The phase diagram of the Creutz ladder model with $\gamma_d$ on the phase boundary being marked. There are two degenerate points in the Brillouin zone for the system with $|M|<2$ and $\alpha=\pi$ or $0$, and the summation of $\gamma_d$ is $\pm 2\pi$. } \label{Fig2}
\end{figure}

Similar scheme can be directly applied to the other 1D topological nontrivial systems with the ground state characterized by the $Z$-type invariant. According to the ten-fold-way classification \cite{tenfold}, in one dimension the BDI (orthognal), AIII (unitary) and CII (symplectic) classes belong to the $Z$ type. As the SSH model belongs to the BDI class, next we consider the fermionic Creutz ladder \cite{Creutz,Bermudez}, which is related to the 1D AIII topological insulator. The Creutz ladder can be
described by the Hamiltonian:
\begin{eqnarray}
H=&&\sum_i Ke^{-i\alpha}\hat{c}_{A,i+1}^{\dagger}\hat{c}_{A,i}+Ke^{i\alpha}\hat{c}_{B,i+1}^{\dagger}\hat{c}_{B,i}+K\hat{c}_{B,i+1}^{\dagger}\hat{c}_{A,i}\nonumber\\
&+&K\hat{c}_{A,i+1}^{\dagger}\hat{c}_{B,i}+M\hat{c}_{A,i}^{\dagger}\hat{c}_{B,i}+h.c.,
\end{eqnarray}
where $K$ and $M$ are tunneling strengths, $\alpha$ is a magnetic flux, and $K=1$ is set to be the energy unit. In the momentum space, the
Hamiltonian $h(k)$ is given by $h(k) = h_0I + h_x\sigma_x + h_z\sigma_z$, with $h_0 = 2 \cos{\alpha}\cos{k}$, $h_x = 2\cos{k}+M$,
$h_z(k) = -2 \sin{\alpha}\sin{k}$. Generally, this model has no time-reversal symmetry due to the introduction of $\alpha$. For the specific case with $\alpha=\pm \pi/2$, one can check that the Hamiltonian fulfills the chiral symmetry $\sigma_y h(k) \sigma_y = - h(k)$, which means the system with $\alpha=\pm \pi/2$ belonging to the AIII class. This model also has a two-band spectrum, and the gap closes when $\sin{\alpha}=0$, $M=-2\cos{k}$ or $M=\pm2$, $\cos{k}=\mp 1$. The phase diagram is shown in Fig.\ref{Fig2}, which indicates the phase within the parameter regime of $|M|<2$ is topologically nontrivial characterized by the Zak phase $\gamma=\pm\pi$, whereas the phase with $|M|>2$ is a trivial phase with $\gamma=0$. Notice that although the Zak phase usually only takes $0$ or $\pi$ with a modulus $2\pi$, the winding directions of $h(k)$ are opposite to each other for $\gamma=\pi$ and $\gamma=-\pi$, which represent different topological phases. It is interesting to point out that the state at $\alpha=0$ and $\pi$ within $|M|<2$ is a 1D semimetal with band touching points at $k=\pm \arccos(-M/2)$. At the phase boundary $M=\pm 2$, which separates the trivial and topological phases, the band gap closes at $k=\pi$ or $0$. Similar to Eq.(\ref{gammad}), we can calculate the Berry phase around the gap closing point at $(M=2, k=\pi)$ or  $(M=-2, k=0)$. As displayed in Fig.\ref{Fig2}, we have $\gamma_d=\pm \pi$ at the phase boundary, which corresponds to the change of the Zak phase $\gamma$ at different phases. If we fix $M$ and take $\alpha$ as a driving parameter, we can calculate the Berry phase around the band touching points $(\alpha=0, k=\pm \arccos(-M/2))$ or $(\alpha=\pi, k=\pm \arccos(-M/2))$. For $\alpha=0$, we get $\gamma_d=-\pi$ at both gap closing points, and for $\alpha=\pi$ we get $\gamma_d=\pi$. It is clear that the summation of $\gamma_d$ for these two degenerate points gives $-2 \pi$ or $2 \pi$, consistent with the change of the Zak phase $\gamma$ at different sides of $\alpha=0$ or $\alpha=\pi$.

We have indicated that the Creutz ladder system with $\alpha=\pm \pi/2$ belongs to the AIII class. When $\alpha$ deviates $\pm \pi/2$, the chiral symmetry is broken and the system can not be classified into the standard ten-fold classes \cite{tenfold}. Nevertheless, the system still supports topologically nontrivial phase with the Zak phase $\gamma=\pm\pi$ as the system is protected by the inversion symmetry \cite{Zak,Inversion}, i.e., $\sigma_x h(k) \sigma_x = h(-k)$. Through the above examples, it is clear that our scheme works for both the standard topological classes, e.g., the BDI class and the AIII class, and the non-standard class protected by other symmetries as long as the topological phase can be characterized by a nontrivial Berry phase.

\subsection{2D topological models}
Next we apply a similar scheme to study the topological property of phase transition points of 2D
systems. We begin our discussion with the famous Haldane
model \cite{Haldane}, which supports a rich phase diagram,
exhibiting either topological or trivial phase transitions. The Haldane model is a prototype model which
may realize the anomalous quantum Hall effect in a 2D honeycomb
lattice without any net magnetic flux through a unit cell of the
system. The Hamiltonian of the Haldane model is given by:
\begin{eqnarray}
H=\sum_i t_0\hat{c}_i^{\dagger}\hat{c}_i+\sum_{\langle i,j\rangle}
t_1\hat{c}_i^{\dagger}\hat{c}_j+\sum_{\langle\langle
i,j\rangle\rangle} t_2e^{i\alpha_{i,j}}\hat{c}_i^{\dagger}\hat{c}_j,
\end{eqnarray}
where the summation is defined on the 2D honeycomb lattice, which is
composed of two sublattices labeled by A and B, respectively. Here
$t_0=M$ for site A and $t_0=-M$ for site B, $t_1$ denotes the
nearest-neighbor hopping amplitude, and $t_2$ denotes the next-nearest-neighbor (NNN)
hopping amplitude. The magnitude of the phase is set to be
$|\alpha_{i,j}|=\alpha$, and the direction of the positive phase is
clockwise, following Haldane's work. This model is well known for
its three topologically different phases characterized by the
Chern number $C$ ($C=\pm 1$ or $0$), which is defined as the integral of the Berry
curvature $\textbf{V}$ for each band in the Brillouin zone, with
$\textbf{V}$ defined as
$\nabla\times i \langle\varphi(\textbf{k})|\nabla|\varphi(\textbf{k})\rangle$, where $\varphi(\textbf{k})=\varphi(k_x,k_y)$
is the eigenstate of the occupied Bloch band.
\begin{figure}
\includegraphics[width=1\linewidth]{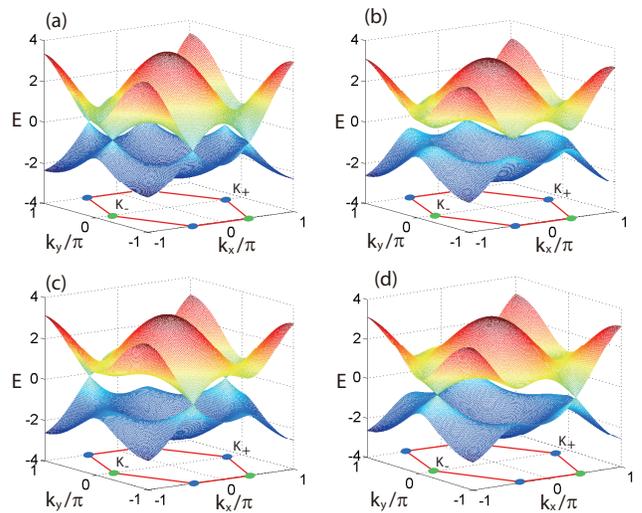}
\caption{(Color online) The energy spectrum of the Haldane model with $t_1=1$ and
$t_2=0.1$. (a) $M=0$, $\alpha=0$; (b) $M=0$, $\alpha=\pi/3$; (c)
$M=3\sqrt{3}t_2\sin{\alpha}$, $\alpha=\pi/3$; (d)
$M=-3\sqrt{3}t_2\sin{\alpha}$, $\alpha=\pi/3$. (a), (c) and (d) are
on the phase boundary, while (b) is in the gap opened region. The
gap closes at both Dirac points in (a), while it closes only at one
of the two Dirac points in (c) and (d), respectively.} \label{Fig3}
\end{figure}

Consider the case with $|t_2/t_1|< 1/3$, for which the two bands never overlap and only touch at the the Brillouin zone corner when $M=\mp3\sqrt{3}t_2\sin{\alpha}$. Expanding the Hamiltonian in the momentum space around the Dirac point
$\textbf{K}_{\pm}=(\pm\frac{4\pi}{3\sqrt{3}},0)$, i.e.
$k_x=\pm\frac{4\pi}{3\sqrt{3}}+x$ and $k_y=y$, we get the effective Hamiltonian:
\begin{eqnarray}
h(\textbf{k})=\mp
\sqrt{3}t_1x\sigma_x+t_1y\sigma_y+(M\mp3\sqrt{3}t_2\sin{\alpha})\sigma_z .
\end{eqnarray}
The gap closing points are at $x=y=0$ and
$M\mp3\sqrt{3}t_2\sin{\alpha}=0$. While the gap closes at both of $\textbf{K}_{\pm}$
when $M=0$ and $\alpha=0$ or $\pi$, there can be no more than
one gap closing point for other value of $\alpha$, i.e., the gap closes at $\textbf{K}_{+}$ when $M=3\sqrt{3}t_2\sin{\alpha}$ or at $\textbf{K}_{-}$ when $M=-3\sqrt{3}t_2\sin{\alpha}$. Different
situations of band touching are shown in Fig.\ref{Fig3}.
\begin{figure}
\includegraphics[width=1\linewidth]{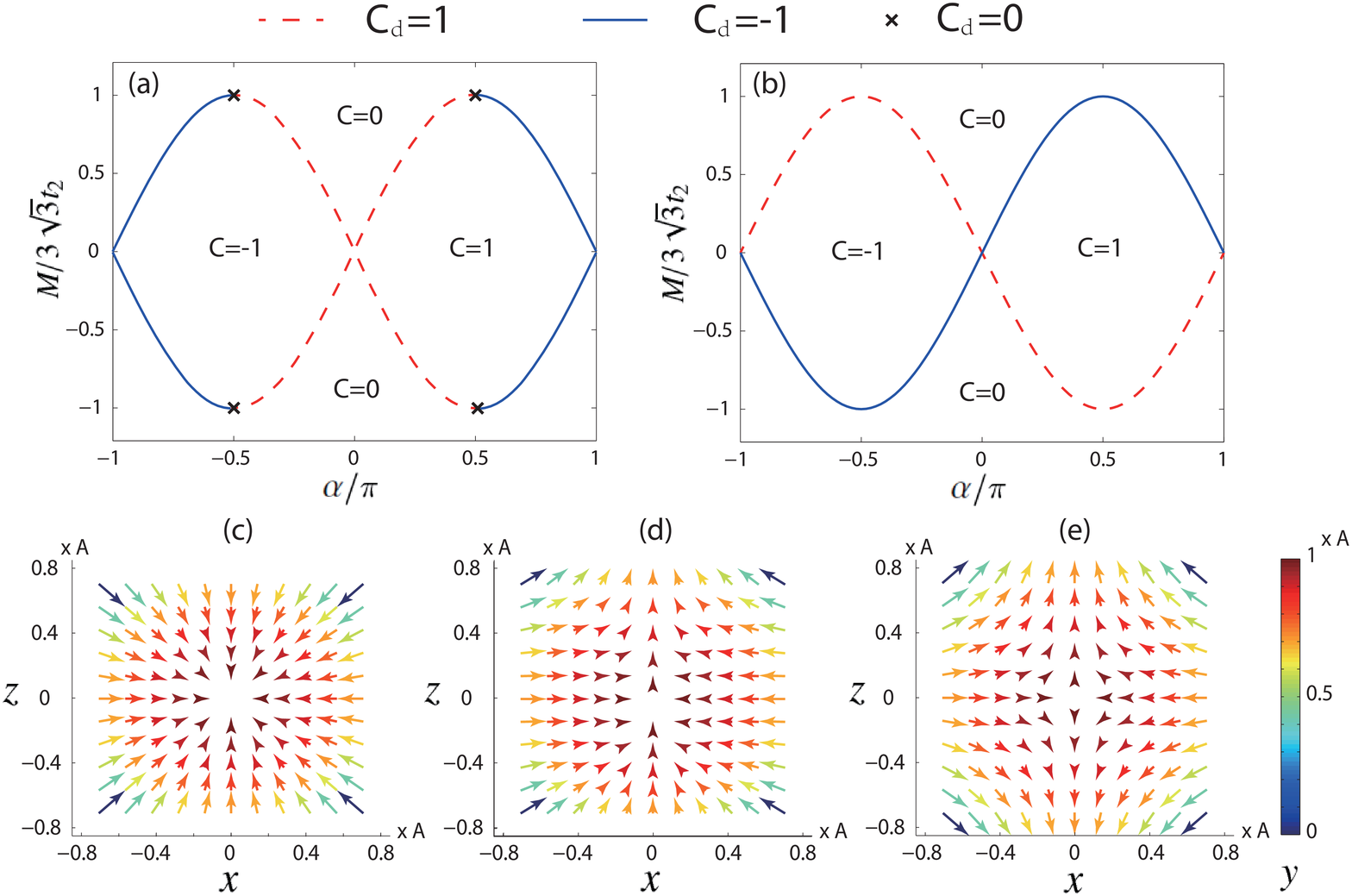}
\caption{ (Color online) (a) and (b) show the phase diagram of the Haldane model with $C_d$ on the phase boundary being marked.
 (a) is for the case with $C_d$ defined in the space of
momentum and $\alpha$; (b) is for the case with $C_d$ defined in the
space of momentum and $M$. (c)-(e) show the direction of
$h(\textbf{k})$ near the gap closing point in (a) with different $\alpha_0$:
(c) $\alpha_0=\pi/3$, $M=9t_2/2$, $C_d =1$; (d)
$\alpha_0=\pi/2$, $M=3\sqrt{3}t_2$, $C_d=0$ ; (e) $\alpha_0=2\pi/3$,
$M=9t_2/2$,  $C_d =-1$. } \label{Fig4}
\end{figure}

The phase diagram of Haldane model is displayed in  Fig.\ref{Fig4} (a) or (b) with different phases characterized by the Chern number $C$. At the phase
boundary, the gap between the upper and lower energy bands closes, and the Chern number is ill-defined therein.
Similar to the 1D case, we can define a topological invariant on a closed surface surrounding the phase transition point in the
parameter space of $\textbf{k}$ and transition driving parameter $\delta$ to describe the topological property
of the transition point.
Different from the 1D system, here the topological invariant is given by
\begin{eqnarray}
C_d=-\frac{1}{2\pi} \oiint \textbf{V}\cdot d\textbf{S},
\end{eqnarray}
with $\textbf{S}$ a sphere surface around the gap closing point
$(k_x^0,k_y^0,\delta_0)$. Following Berry's work \cite{Berry}, the
Berry curvature for the lower band can be expressed as:
\begin{eqnarray}
\textbf{V}=Im\frac{\langle-|\nabla h(\textbf{k},\delta) |+\rangle\times\langle+|\nabla
h(\textbf{k},\delta) |-\rangle}{(E_--E_+)^2},
\end{eqnarray}
with $|+\rangle$ ($|-\rangle$) the eigenstate of the upper (lower)
band. With some further calculations, the Berry curvature can be
written as:
\begin{eqnarray}
V^x &=&-\frac{1}{2R^3}[\frac{\partial h_x}{\partial
k_y}\frac{\partial h_y}{\partial \delta}h_z+\frac{\partial
h_y}{\partial k_y}\frac{\partial h_z}{\partial
\delta}h_x+\frac{\partial h_z}{\partial k_y}\frac{\partial
h_x}{\partial
\delta}h_y\nonumber\\
&&-\frac{\partial h_y}{\partial k_y}\frac{\partial h_x}{\partial
\delta}h_z-\frac{\partial h_x}{\partial k_y}\frac{\partial
h_z}{\partial \delta}h_y-\frac{\partial h_z}{\partial
k_y}\frac{\partial h_y}{\partial
\delta}h_x],\nonumber\\
V^y&=&-\frac{1}{2R^3}[\frac{\partial h_x}{\partial
\delta}\frac{\partial h_y}{\partial k_x}h_z+\frac{\partial
h_y}{\partial \delta}\frac{\partial h_z}{\partial
k_x}h_x+\frac{\partial h_z}{\partial \delta}\frac{\partial
h_x}{\partial k_x}h_y\nonumber\\
&&-\frac{\partial h_y}{\partial z}\frac{\partial h_x}{\partial
k_x}h_z-\frac{\partial h_x}{\partial z}\frac{\partial h_z}{\partial
k_x}h_y-\frac{\partial h_z}{\partial z}\frac{\partial h_y}{\partial
k_x}h_x],\nonumber\\
V^z&=&-\frac{1}{2R^3}[\frac{\partial h_x}{\partial
k_x}\frac{\partial h_y}{\partial k_y}h_z+\frac{\partial
h_y}{\partial k_x}\frac{\partial h_z}{\partial
k_y}h_x+\frac{\partial h_z}{\partial k_x}\frac{\partial
h_x}{\partial k_y}h_y\nonumber\\
&&-\frac{\partial h_y}{\partial k_x}\frac{\partial h_x}{\partial
k_y}h_z-\frac{\partial h_x}{\partial k_x}\frac{\partial
h_z}{\partial k_y}h_y-\frac{\partial h_z}{\partial
k_x}\frac{\partial h_y}{\partial k_y}h_x].\nonumber
\end{eqnarray}
Here the Hamiltonian
$h(\textbf{k})=h_x\sigma_x+h_y\sigma_y+h_z\sigma_z$ and
$R=\sqrt{h_x^2+h_y^2+h_z^2}$. For convenience, we choose a spherical surface
surrounding the gap closing point with a radius of $A$, i.e.,
$k_x=k_x^0+x,~k_y=k_y^0+y,~\delta=\delta_0+z$ with
$x=A\sin{\theta}\cos{\phi}$, $y=A\sin{\theta}\sin{\phi}$ and
$z=A\cos{\theta}$, where $\theta$ is the polar angle and $\phi$ is the
azimuthal angle of the spherical surface. The integral then becomes:
\begin{eqnarray}
C_d=-\frac{1}{2\pi}\oiint
\textbf{V}\cdot(\sin{\theta}\cos{\phi},\sin{\theta}\sin{\phi},\cos{\theta})
A^2\sin{\theta}d\theta d\phi.\nonumber
\end{eqnarray}
If all the $h_x$, $h_y$ and $h_z$ are linear of $(x,y,z)$ around a
gap closing point, we can rotate and stretch the axes to reach
$h_x= \pm x,~h_y= \pm y,~h_z= \pm z$, and the Chern number $C_d=\pm
1$.

As the phase boundary is given by $|M| =3\sqrt{3}|t_2\sin{\alpha}|$, either the parameter $\alpha$
or $M$ can be chosen as the phase transition driving parameter. Similar to the 1D cases, we can judge whether the phase transition is
topological  or not by examining $C_d$ defined around the gap closed points.
First we choose $\alpha$ as the third parameter besides the momentum
$\textbf{k}$ by keeping $M$ fixed. We take the case with the transition point located at $\textbf{K}_+$ as an
example, and the case at $\textbf{K}_-$ can be analyzed similarly. Defining
$\alpha=\alpha_0+z$ with $M-3\sqrt{3}t_2\sin{\alpha_0}=0$ and expanding
the Hamiltonian near $\alpha_0$, we have
\begin{eqnarray}
h(\textbf{k})=-
\sqrt{3}t_1x\sigma_x+t_1y\sigma_y-3\sqrt{3}t_2\cos{\alpha_0}z\sigma_z
\label{hk1}
\end{eqnarray}
when $|M|<3\sqrt{3}|t_2|$, and the Chern number around
$\textbf{K}_+$ is $C_d(\alpha)=\text{sgn} (\cos{\alpha_0})$. When
$|M|=3\sqrt{3}|t_2|$, i.e., the case with $\alpha_0=\pm \pi/2$
marked by the ``cross" in Fig.\ref{Fig4}(a),  however, the expansion
becomes
\begin{eqnarray}
h(\textbf{k})=-
\sqrt{3}t_1x\sigma_x+t_1y\sigma_y+3\sqrt{3}t_2z^2\sigma_z,\label{hk2}
\end{eqnarray}
and the integral results in $C_d(\alpha)=0$ as the integrand is
an odd function in the interval. In
Fig.\ref{Fig4}(a), we show the value of $C_d$ around the phase boundary: except $C_d=0$ at the four points marked by  ``crosses", it takes either $1$ or $-1$. We can see that $C_d(\alpha)$ shows the
change of the band Chern number $C$ across the transition point by varying $\alpha$.
For the case of $C_d =0$, the change of $C$ is zero when varying
$\alpha$, which indicates a topologically trivial phase transition.
On the other hand, the case of $C_d = \pm 1$ corresponds to a topological phase transition from the trivial (topological) phase to topological (trivial) phase.
Particularly, when $M=0$ and $\alpha=0$ (or $\pi$), there are two gap closing points in the Brillouin zone,
and the change of $C$ is $\pm2$ when varying $\alpha$,  which corresponds to the summation
of $C_d$ around these two points.

If we choose $M$ as the third parameter besides the momentum
$\textbf{k}$ by keeping $\alpha$ fixed and define $M=M_0+z$ with
$M_0\mp3\sqrt{3}\sin{\alpha}=0$, the expansion of the Hamiltonian
becomes:
\begin{eqnarray}
h(\textbf{k})=\mp \sqrt{3}t_1x\sigma_x+t_1y\sigma_y+z\sigma_z,
\end{eqnarray}
and the Chern number around the gap closing point of $\textbf{K}_+
(\textbf{K}_-)$ is always $C_d(M)=-1(1)$, as shown in
Fig.\ref{Fig4}(b). The value of $C_d(M)$ also indicates the change
of $C$ when varying $M$. When $M=0$ and $\alpha=0$ and $\pi$, the
change of $C$ is zero, which also matches the summation of $C_d(M)$.

%From Eq. (\ref{hk1}) and (\ref{hk2}) we
%notice that the difference of $C_d$ for Haldane model comes from the
%coefficient of $\sigma_z$.
To visualize the Chern number $C_d$, we show the direction of
$h(\textbf{k})$ around the gap closing point with different $\alpha_0$
in Fig.\ref{Fig4}(c)-(e). One can see that all the
arrows point to $z=0$ in (c), and to the opposite direction in (e),
but the z-component of $h(\textbf{k})$ in (d) is always positive.
The value of $C_d$ is associated with the direction of
$h(\textbf{k})$. In Fig.\ref{Fig4}(c), the direction of
$h(\textbf{k})$ is always toward $x=z=0$ and away from $y=0$, which
corresponds $C_d=1$. In Fig.\ref{Fig4}(e), the direction of
$h(\textbf{k})$ is opposite to the one in Fig.\ref{Fig4}(c) only in
the $z$ direction, corresponding to $C_d=-1$.
%In
%Fig.\ref{Fig4}(d), however, the direction of $h(\textbf{k})$ varies
%in the $z$ direction, which correspond to $C_d=0$.
%One can see that all the arrows point to $z=0$ in (c), and to the
%opposite direction in (e), which correspond to $C_d=-1$ and $C_d=1$
%respectively.
However, in Fig.\ref{Fig4}(d), the $z$-component of
$h(\textbf{k})$ follows the same direction, corresponding to $C_d=0$.
\begin{figure}
\includegraphics[width=1\linewidth]{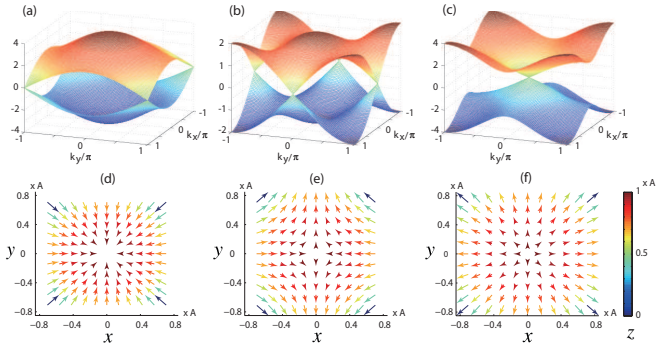}
\caption{(Color online) (a)-(c) The energy spectrum of the QWZ model with
$\beta=1$, (a) $\mu=-2$, (b) $\mu=0$ and (c) $\mu=2$, respectively. (d)-(f) display
the direction of the Hamiltonian $\textbf{h}$ in the in the x-y
plane, corresponding to the gap closing points in (a)-(c)
respectively. In (e) we only display the direction of Hamiltonian
around one of gap closing points at ($k_x=\pi$, $k_y=0$) shown in (b), while
the other one at ($k_x=0$, $k_y=\pi$) has exactly opposite direction in
both the $x$ and $y$ direction. } \label{Fig5}
\end{figure}

Our theory can be applied to study other 2D Z-type topological systems with the
topological invariant characterized by the Chern number.
To give an additional example, we next investigate a lattice model on a
square lattice described by the Hamiltonian
$H=\mathbf{h}\cdot\mathbf{\sigma}$ with
$\mathbf{\sigma}=(\sigma_x,\sigma_y,\sigma_z)$ the Pauli matrices
and
$\textbf{h}=(\sin{k_x},\sin{k_y},\beta[\mu-\cos{k_x}-\cos{k_y}])$, which
 is known as the Qi-Wu-Zhang (QWZ) model \cite{QWZ}.
%of the quantum anomalous Hall effect.
Depending on the value of $\mu$, this model
has three different topological phases characterised by the band
Chern number $C$. Setting $\beta=1$, the band Chern number $C=0$
while $|\mu|>2$, and $C=-\text{sgn}(\mu)$ when $|\mu|<|2|$. There are three
phase transition points for this model: $\mu=-2$, $\mu=0$ and
$\mu=2$. As shown in Fig.\ref{Fig5}(a)-(c), when $\mu=\pm2$,
there's only one gap closing point ($k_x=k_y=\pi$ for $\mu=-2$ and
$k_x=k_y=0$ for $\mu=2$) in the Brillouin zone, but when $\mu=0$,
there are two gap closing points ($k_x=\pi,~k_y=0$ and
$k_x=0,~k_y=\pi$). By choosing $\mu$ as the third parameter beside the
momentum, the expansion of the Hamiltonian near the gap closing
points results in
\begin{eqnarray}
h(\textbf{k})&=&-x\sigma_x-y\sigma_y+z\sigma_z,~\mu=-2; \nonumber \\
h(\textbf{k})&=&\pm x\sigma_x\mp y\sigma_y+z\sigma_z,~\mu=0; \nonumber \\
h(\textbf{k})&=&x\sigma_x+y\sigma_y+z\sigma_z,~\mu=2. \nonumber
\end{eqnarray}
In Fig.\ref{Fig5}(d)-(f) we show the direction of $\textbf{h}$ in
the $x-y$ plane, as the $z$ direction of $h(\textbf{k})$ is always
away from $z=0$. The result shows that $C_d=1$ for $\mu=\pm2$, and
$C_d=-1$ for both gap closing points when $\mu=0$, hence the
summation of $C_d$ at each phase transition point also corresponds
to the the change of the band Chern number $C$.
%Similar to the 1D cases, our scheme can be also applied to study

\section{Summary and outlook}
In summary, we proposed a scheme to study the topological properties of phase transition point for various topological QPTs by introducing a topological invariant defined on a closed curve in the parameter space surrounding the transition point. By studying several typical topological models, we demonstrated that the topological or trivial phase transition can be distinguished by the introduced topological invariant around the transition point, which takes nontrivial quantized numbers for topological QPTs but non-universal numbers or zero number for conventional QPTs. Our theory provides a way to discriminate topological or trivial QPTs by directly studying the properties of the phase transition point.

In order to give a proper definition to the topological invariant of a phase transition point, we have introduced the phase transition driving parameter as
an additional dimension parameter besides the momentum space, and thus raise the effective parameter dimension
by one. Such an idea seems to share some similarities with the construction in the analysis of topological phases by dimension extension through introducing an additional dynamical parameter \cite{Lang,Xu,ZhangFan,Prodan}. While the topological invariant is defined in the enlarged $(D+1)$-dimensional parameter space for $D$-dimensional systems in previous works \cite{Lang,Xu,ZhangFan,Prodan}, in the present work the topological invariant is defined on a close curve of $(D+1)$-dimensional space surrounding the phase transition point and thus is effectively defined on a $D$-dimensional curve.  This is the main difference from the previous construction via dimension extension. While our scheme can be applied to $Z$-type topological systems, it can not be directly generalized to deal with the $Z_2$-type systems as the corresponding topological invariant is described by the $Z_2$ number \cite{tenfold}. It would be interesting to study how to characterize the topological phase transition for $Z_2$-type systems by analysing the topological property of the phase transition point in the future work.

\begin{acknowledgments}
%%%%%%%%%%%%%%%%%%
S. C. would like to thank H. M. Weng for helpful discussions. The work is supported by NSFC under Grants No. 11425419, No. 11374354 and No. 11174360.
\end{acknowledgments}

\end{document}